\documentclass[twocolumn,10 pt,showpacs,preprintnumbers,amsmath,amssymb]{revtex4}
\usepackage{amssymb}
\usepackage{graphicx}
\begin{document}

\title{Contrasted role of disorder for magnetic properties in an original mixed valency iron Phosphate.}
\author{Laure Adam, Alain Pautrat, Olivier Perez and Philippe Boullay}
\affiliation{Laboratoire CRISMAT, UMR 6508 du CNRS, ENSICAEN et Universit\'{e} de Caen, 6 Bd Mar\'{e}chal Juin, F-14050 Caen 4.}

\begin{abstract}                                                                                                                                                                            

We have measured the magnetic properties of a mixed valency iron phosphate.
 It presents an original structure with crossed chains containing $Fe^{II}$ and orthogonal to the longest direction of the crystallites.
 Microstructural investigations using electron microscopy show the presence of random nano-twinning.
 The ac susceptibility measurements demonstrate similarities with the kinetics of a disordered magnetic, spin-glass like, state but
 are shown to be essentially due to this peculiar disorder.
 Scaling properties are characteristics of 3D second order transition implying that this disorder at a small scale
 does not influence significantly long range magnetic ordering.
At low temperature, a decrease of the spontaneous magnetization and an irreversible metamagnetic transition is observed,
and is attributed to a canting of the spins in the iron chain.

\end{abstract}

\newpage
\maketitle

\section{Introduction}

Transition metal phosphates constitute a great family of compounds widely
 investigated because of their various potential applications, such as cathodes for Li batteries \cite{good},
 heterogeneous catalysts \cite{cata} ,
   or molecular sieves \cite{mole}. Magnetic properties are less deeply studied although structural
 features of many reported phases  
(e.g. metal polymers, infinite chains of metal polyhedra) 
suggest interesting, albeit complex, magnetic interactions. Indeed, the structural features mixes low dimensionality,
 disorder and, more rarely, different valencies which may lead to different spin sublattices.
Up to now, vanadium phosphates still seem to be the most investigated \cite{soso,vana1,vana2}. 
Others transition metal phosphates are obviously studied such as copper phosphates \cite{copper1,copper2}, or manganese
 phosphates for instance \cite{manga}.
 Focusing on the iron phosphate group, an antiferromagnetic behaviour is reported for $Fe_2P_2O_7$ (superexchange mechanism between $Fe^{II}$) \cite{Fe1},
 $Fe_3PO_7$ (direct exchange or superexchange mechanisms between $Fe^{III}$) \cite{Fe2} and $Fe_4(P_2O_7)_3$
 (superexchange mechanism between $Fe^{III}$) \cite{Fe3}.
 $SrFe_2(PO_4)_2$, which framework exhibits iron tetramers in the high spin $Fe^{II}$ state, present more complex magnetic properties,
 including metamagnetic regimes and
 competition between structural and magnetic phase transitions \cite{Fe4,Fe5}. A study by powder neutron
 diffraction of $\beta-Fe_2(PO_4)O$,
 which 3D-framework is built of iron polyhedra infinite chains, revealed a ferromagnetic order within the chains and an antiferromagnetic order
 between the chains which leads to an antiferromagnetic bulk transition \cite{Fe6}.
 One interesting aspect of transition metal phosphate for magnetism is that phosphate groups can isolate magnetic ions,
 leading to low dimensional spin systems. Another interesting aspect, maybe less deeply studied, is that
 original structures with a non trivial disorder can be stabilized. As we will show here, 
such a peculiar disorder can have important consequences with for example an averaging of low dimensional magnetic characteristics at the sample scale.
 Then, coupling a careful analysis of structural data with detailed magnetic measurements will help to shed some light on 
the role of specific geometries for magnetic properties of disordered medium.

 We report herein on the study of magnetic properties in the original mixed valency iron phosphate $\beta-(NH_4)Fe_2(PO_4)_2$
 using macroscopic magnetic measurements. 
The study by X-Ray diffraction of its structure has already been reported \cite{cristallo} so that only a brief reminder of the
 structural main features will be given. An additional study using electron microscopy allows a better understanding of the
 disorder phenomenon at a local scale and its influence on the magnetic properties, which reveal interesting features such as slow dynamics
 due to the peculiar disorder of the structure and canting of the spins at low temperature.

\section{sample preparation and structural analysis}

A single-phase sample of $\beta -(NH_4)Fe_2(PO_4)_2$ was prepared under hydrothermal conditions.
 A mixture of $Fe_2O_3$, $FeCl_2$ and $(NH_4)_2HPO_4$, in the respective molar ratio 0.25:0.5:1,
 was placed in a 21 mL Teflon-lined stainless steel autoclave with 2 mL of deionized water.
 It was heated at 493 K over 25 H and then cooled to room temperature for 17 H.
 The resulting product, made up of black needle-like crystals of $\beta -(NH_4)Fe_2(PO_4)_2$ (Fig.1),
 was filtered and washed with deionized water.
The phosphate $\beta -(NH_4)Fe_2(PO_4)_2$ was then characterized using single-crystal X-Ray diffraction,
 and crystallizes in the orthorhombic space group $F_{ddd}$ with the cell parameters a = 17.1491(2) \AA, b = 7.4419(2) \AA, c = 10.0522(3) \AA.
 Detailed characterization using superspace formalism was previously proposed \cite{cristallo}. In particular, the presence
 of diffuse scattering characteristic of correlated disorder was reported. The three-dimensional framework of $\beta -(NH_4)Fe_2(PO_4)_2$ is built of distorted $Fe^{II}O_6$ and $Fe^{III}O_6$ octahedra and $PO_4$ tetrahedra.
 Theses polyhedra share edges and apices to form hexagonal tunnels, running along $[011]$ and $[0\bar{1} 1]$ directions,
 in which ammonium cations are located (Fig.2).
Projections of the structure along $[011]$ and $[0\bar{1}1]$ directions show that $[FeP_2O_8]_\infty$ ribbons are running along $[011]$ and $[0\bar{1}1]$ 
directions (Fig.2), i.e. perpendicularly to the needle-like crystal axis $\vec{a}$; the angle between two ribbons being
$73^{\circ}$ (Fig.3). Each $[FeP_2O_8]_\infty$ ribbon is made of edge-sharing $Fe^{II}O_6$ and $PO_4$ polyhedra; the role of the 
$[FeO_4]_\infty$ chains (Fig.2) will be discussed in the analysis of the magnetic properties. $[FeP_2O_8]_\infty$ ribbons are connected to 
each other thanks to $Fe^{II}O_6$ octahedra and $PO_4$ tetrahedra. As a consequence, one $Fe^{II}O_6$ octahedron share one edge 
with one $PO_4$ tetrahedron, four apices with four other $PO_4$ tetrahedra and two apices with two $Fe^{III}O_6$ octahedra.
The $Fe^{III}O_6$ octahedra running along $[011]$ and $[0\bar{1}1]$ directions, i.e. the ordered part of this complex 
structure, form hexagonal tunnels containing $NH^{4+}$
 cations alternate with $[FeO_4]_8$ chains according to the sequence $NH^{4+}-[FeO_4]_8-NH^{4+}-[FeO_4]_8$ (Fig.2).
 Nevertheless, a random distribution of hexagonal tunnels and $[FeO_4]_8$ chains can be observed along  the $\vec{w}$  direction,
 \textit{i.e.} $[2\bar{1}1]$.
 Such an impossibility to define a fixed sequence along this direction indicates that $\beta -(NH_4)Fe_2(PO_4)_2$ 
exhibits a complex disorder phenomenon along $\vec{a}$.

\section{Microstructural analysis using transmission electron microscopy}

In order to obtain further information on the nature of this disorder at a local scale, we decided to investigate
 $\beta-(NH_4)Fe_2(PO_4)_2$ by Transmission Electron Microscopy (TEM) using a JEOL2010F microscope. 
Part of the product of the hydrothermal synthesis was crushed in an agate mortar to obtain small fragments 
that were put in a suspension in alcohol. A drop of the suspension was then deposited and dried on a copper grid
 previously coated with a thin film of amorphous carbon. Selected Area Electron Diffraction (SAED) performed on several crystallites confirms that  
%As shown from the single crystal X-ray diffraction study \cite{cristallo},
 $\beta$-(NH$_4$)Fe$_2$(PO$_4$)$_2$ presents a disordered structure evidenced on the diffraction patterns
 by the systematic presence of diffuse scattering along the [100]* direction.  

As mentioned above, the $[011]$ projection (Fig.2) appears the one to choose to obtain
 information on how the hexagonal tunnels are filled: either with NH$_4^+$ groups or with [FeP$_2$O$_{4}$]$_{\infty}$
 ribbons. In the Fig.4a, a typical $[011]$ SAED zone axis patterns (ZAP) shows spots that can be
 indexed considering the $F_{ddd}$ orthorhombic unit cell previously reported \cite{cristallo} (Fig.4b).
 The Fig.4a also reveals the existence of strong diffuse streaks (marked with white arrows) parallel
 to [100]* and located in between the spots rows. In addition, depending on the crystallite under observation,
 more or less pronounced intensity reinforcement in the form of nodes elongated along [100]* can be observed
 at positions that could be indexed considering a slightly distorted [112] ZAP of the $\alpha$-(NH$_4$)Fe$_2$(PO$_4$)$_2$
 monoclinic unit cell \cite{alpha-sosso} (Fig.4c). 
%This information already let us imagine that the disordered stacking sequence along {\bf a} contains blocks of limited thickness related to $\alpha$-(NH$_4$)Fe$_2$(PO$_4$)$_2$. 
Keeping in mind that our goal was to reveal how the [FeP$_2$O$_{4}$]$_{\infty}$ ribbons are locally
 arranged in the structure and also considering that this compound was beam sensitive, we choose
 to gain information at a local scale working at a medium resolution in the TEM imaging mode. 
% using a relatively small objective aperture. 
Using an objective aperture to select diffraction scattered at an angle corresponding to a $\sim$5\AA\ resolution limit
 (see light gray circle in Fig.4a), we form an image with only few beams but with the most intense
 part of the diffuse scattering. The structure of the ordered $\alpha$-(NH$_4$)Fe$_2$(PO$_4$)$_2$ was used to
 simulate the TEM images (JEMS software) and verify that in such an operating condition the images actually allow
 exclusively to reveal the arrangement of the hexagonal tunnels (separated at least by 6\AA). Additionally there
 is a large focal range where the hexagonal tunnels filled with NH$_4^+$ groups appear as bright intense dots for
 thicknesses above 10nm. The experimental image Fig.5 shows a crystallite heavily twinned at the
 nanoscale with a twin plane perpendicular to the [100] direction. 
The Fig.5a is a Fourier filtered image constructed by selecting only the six spots present in the
 Fourier diffractogram of the experimental image and operating an inverse Fourier transformation. Such an image
 actually represents the arrangement of the hexagonal tunnels regardless their occupancy (see schematic representation
 above the Fig.5a). It do not exhibit any twinning and the periodicity along $[0\bar{1}1]$ is half the
 one observed in the experimental image. 
It shows that the arrangement of the hexagonal tunnels is identical all over the experimental image, only the way
 they are filled up differs. Importantly, this last information is carried on by the diffuse streaks and can be related, 
in the image mode, to the observed twinning. 
The enlarged parts of the experimental image Fig.5b and c show two different types of nano-structure.
 The first one Fig.5b corresponds to a twin structure where the hexagonal tunnels are filled following 
the sequence NH$_4^+$-[Fe$P_2$O$_{4}$]$_{\infty}$-NH$_4^+$-[Fe$P_2$O$_{4}$]$_{\infty}$ along the direction {\bf w}$\sim$[2$\bar{1}$1].
 This sequence is related to the stacking found in the ordered $\alpha$-(NH$_4$)Fe$_2$(PO$_4$)$_2$. 
The second one Fig.5c corresponds locally to a structure where the hexagonal tunnels are
 filled following the sequence NH$_4^+$-NH$_4^+$-[Fe$P_2$O$_{4}$]$_{\infty}$-[Fe$P_2$O$_{4}$]$_{\infty}$ along 
the direction {\bf w}, which is related to the ideally ordered $\beta$-(NH$_4$)Fe$_2$(PO$_4$)$_2$ reported in \cite{cristallo}.
 The disordered structure of $\beta$-(NH$_4$)Fe$_2$(PO$_4$)$_2$ explored at this local scale is thus fully consistent
 with the conclusion previously drawn from the single crystal X-ray diffraction. Locally each 
platelet presents a filling sequence in the hexagonal tunnels that
 can be related either to the $\alpha$- type ordering with different twinned domains or the "ideally" $\beta$ - type ordering, which
 can be considered as polytypes of NH$_4$Fe$_2$(PO$_4$)$_2$.
 The arrangement of NH$_4^+$ groups and
 [Fe$P_2$O$_{4}$]$_{\infty}$ ribbons is ordered along the [0$\bar{1}$1] direction with a sequence NH$_4^+$-[Fe$P_2$O$_{4}$]$_{\infty}$
 (see schematic drawing in Fig.5b and c). The diffuse scattering originates from the absence of NH$_4^+$ / [Fe$P_2$O$_{4}$]$_{\infty}$ 
long range ordering along [100] with the intergrowth of polytype structures and formation of twinning at a nanoscale
 in the form of infinite platelets developed along [0$\bar{1}$1]
 and of limited and random extension along [100].

\section{magnetic properties}

\subsection{General characterization with DC magnetization}
 A 1.57 mg amount of randomly oriented crystals was used for studying of magnetic properties.
 Measurements were made in a MPMS-magnetometer and using ACMS option of a PPMS (Quantum-design).
 Note that to check that our samples behave as a powder of randomly distributed crystallites,
 some measurements on a compacted pure powder of the same phosphate were also conducted, and they produced similar results.
First magnetic characterizations were made using measurements of the dc magnetization.
 In Fig.6 is shown the FC-ZFC thermal magnetization with an applied field of $B=0.1 T$.
Different features can be observed. At $T\approx 24 K$, a large increase of the FC magnetization is observed,
 suggesting a ferromagnetic or ferrimagnetic ordering. At lowest temperature, the FC and ZFC curves separate, and show thermal hysteresis.
 A small decrease of the FC magnetization can be observed, below a temperature $T \approx 8K$.    
 A Curie-Weiss plot of the susceptibility $\chi ^{-1}=C /(T- \theta_c)$ gives the following parameters:
 $\theta_c= -33.6 \pm 0.4 K$ showing antiferromagnetic correlations and $\mu_{eff}=7.12\pm 0.06 \mu_B$ per formula unit.
This latter value is close to one $Fe^{II}$ and one $Fe^{III}$ in a high spin state ($\mu_{eff}=7.68 \mu_B$).
This proportion of $Fe^{II}/Fe^{III}$ is in agreement with both the structural resolution and the Mossbauer measurements at room temperature (not shown here).                                                              

The magnetic cycle at $T=2K$ has a clear ferromagnetic-like signature (Fig.7) after the first magnetization curve,
 with a very low coercivity of $H_c\sim 10G$ classifying our material as
 a soft magnet.  
At the maximum of the magnetic field available in our SQUID MPMS, the magnetization is not fully saturated ($M(H=5T)\sim 2.75 \mu_B/f.u$).
 Complementary measurements in a 14T PPMS using the acms option shows that $M_{sat}\sim 3 \mu_B/f.u$ for $H\geq 9T$. This is much lower than the value of
 $9 \mu_B/f.u$ expected for fully polarized $Fe^{II}$ and $Fe^{III}$ in the high spin state. From this, a ferrimagnetic ordering can be suspected.
 Note that the $\chi^{-1}(T)$ curve has a peculiar rounded shape above the transition temperature (inset of Fig.6)
which is also typical of a ferrimagnetic ground state.
 
\subsection{Frequency dependent susceptibility}

The frequency dependent susceptibility is a classical tool to characterized disordered magnetic states. 
For instance, the measure of the shift $\Delta T$ of the susceptibility maximum for different frequencies $f$
 is an usual procedure to evidence glassy-like features. The parameter $\alpha= \Delta T/ T_0 log(2.\pi.f)$ is often
 used as a phenomenological classification
 of disordered magnetic states \cite{mydosh}. We have measured the ac suceptibility with applied frequencies ranging from $f=1 Hz$ to $1 KHz$.
 We observe a shift of the susceptibility maximum (Fig.8), and calculate $\alpha\approx 3.10^{-3}$, which defines apparently
 our transition as a spin-glass one \cite{mydosh,maud}. 
It is however clear from the preceding paragraph that our sample has a ferromagnetic characteristic with a permanent moment,
 and by definition can not be a spin glass. A well known property of spin glasses, the aging effect \cite{eric},
 is not also observed (the relaxation of the magnetization is found
 to be independent on the time spent at a constant temperature before cutting the field).
 To understand the reason of the slow dynamics observed with the ac susceptibility, we have first studied the variation 
of the transition temperature $T_m$ when the applied magnetic field is increased (Fig.9). This variation is non monotonic.
 At low field $H < H^*\approx 300 G$,$T_m$ decreases
 when H increases \cite{notebene}. For $H>H^*$, the opposite field dependency is observed:
 $T_m$ is an increasing function with $H$.
 This latter evolution, and the observed broadening of the transition region, is expected for the critical region 
 of a second order transition, and is then consistent with a conventional ferromagnetic transition. The question is clearly
 what is the cause of the low field behavior ($H < H^*\approx 300 G$) which mimics the slow dynamic of a spin glass around the transition temperature.
 To answer to this question, we have measured the dc magnetization as function of the applied magnetic field $H$ in the same range of temperatures.
As shown in fig.10, as soon as $T<T_c$, a linear part emerges at low magnetic field values. It corresponds to the straight line of demagnetizing field.
 In the particular case of a ferromagnetic-like transition where the magnetic susceptibility goes to infinity at the transition,
 this slope of this line is simply $1/N$ where $N$ is the coefficient of demagnetizing field. Its origin should be purely geometric.
 Fitting this line provides an estimate of $1/N \approx 2.5$ and $N \approx 0.4$.
 This value is roughly consistent with crystallites of random orientation with respect to the applied field (here, the crystallites have
 approximatively elliptical shapes with a ratio length/diameter of 5 in average).    
 Note the line of demagnetizing field is observed here in a polycrystalline mixture. It shows that that the elementary crystal has not
 a purely uniaxial magnetic symmetry and should be more isotropic \cite{tremolet}, at least at the temperature of the ordering transition.
 Since the crystallites are needle-like, a quasi-1D character could be expected. 
An originality of the structure is that the chains of $Fe^{II}$ are not along the longest dimension of the crystallites, but
 in the plane which is orthogonal to it. As described above, these chains are also along two different directions.
It is then likely that the magnetocrystalline energy favors the orientation of magnetic moments along these iron chains, giving
 an isotropic character in the spins direction.

 From this part, it can be concluded that the (field, temperature) range, where the slow dynamics of spin-glass type
 appear, is directly linked to the demagnetizing effect close to the transition. It is then directly linked
 with the presence and the growing of magnetic domains.
 In some aspects, it recall the Hopkinson 
effect of ferromagnetic which is observed in the low field limit \cite{hop,wyl}, but the frequency dependence
 that we observe indicate additional processes. The small frequency dependence is a 
characteristic of a slow dynamic, and implies here the presence of effective pinning centers which
 impede the growth and nucleation of magnetic domains. We have shown in the structural study that twin boundaries exist parallel
 to the iron chains, and can be responsible for local strains along which domain walls could corrugate.
 The size of the defects is few nanometers, which is a typical scale for domain walls. These extended defects
 are likely responsible
 for domains wall pinning and for the associated slow dynamics. Interaction between twin boundaries and domain walls has
 been already reported in oxydes \cite{TBpinning},
For example, domain walls were observed to be constraint by the twin direction using magneto optical imaging \cite{kees}.
 
 We have proposed potential defects which are important for impeding the growth and nucleation of magnetic domains
 in the magnetization process, but they
 do not contribute to a large coercitive field. Indeed, our sample is a soft magnet with a coercitive field $H_c\approx 20 G$ at $T=2K$.
We think that, as in amorphous material, the local disorder causes fluctuations of the anisotropy from one magnetic
 site to the other and the effective pinning potential for domains wall which are already formed is low.
A consequence can be the existence of 3D rather than low dimensional magnetic interactions at the sample scale,
 what can be studied with the analysis of critical exponents close to the ordering temperature.

\subsection{extraction of critical exponents and Bloch law}

We have shown that the low field regime $H<H^*$ is dominated by demagnetizing and pinning processes. Then, the genuine transition should be 
studied for $H>H^*$ when the variation of the apparent temperature of transition shows signs of a critical behavior (a decrease/broadening of the peak which
 moves to highest temperatures when the magnetic field is increased). This has been already emphasized by different authors,
 who have noted that the scaling laws of critical transition should be studied outside the range where technical processes dominate.
 Following \cite{Williams}, the position and value of the maximum of the ac susceptibility $\chi$ can be related to critical exponents as
 deduced from standard scaling theory. Using the reduced variables $t_m=T_m/T_c-1$ ($T_m$ is the location of the susceptibility maximum) and $h=H_i/T_c$ where
 $H_i$ is the internal field corrected from demagnetizing effect ($H_i=H-NM$), the following power-law dependences should be fulfilled:

\begin{equation}
\chi(H_i,t_m) \propto H_i^{(1/ \delta )-1}
\end{equation}

\begin{equation}
t_m \propto H_i^{1/( \gamma + \beta )}
\end{equation}

\begin{equation}
\chi_m \propto t_m^{- \gamma }
\end{equation}

As shown in Fig.11 and 12, the following values of critical exponents are found: $\delta=4.67 \pm 0.07$, $\gamma=1.47 \pm 0.03$ and $\beta=0.41 \pm 0.06$.
 These values
 are very close to those expected for a 3D Heisenberg isotropic magnet ($\delta=4.80, \gamma=1.39 \pm 0.09$ and $\beta=0.36$) \cite{heisen},
 taken into account the experimental precision.
  This analysis confirms that interactions are long range and essentially 3D,
 and that random disorder of the structure does not impede significantly the long range magnetic ordering.

In conventional ferromagnets (above technical saturation),
 but also when collinear ferrimagnetic structure is achieved at high enough magnetic field , the 
saturation magnetization $M_s$ generally follows the Bloch law derived with spin waves theory. To first order,
 the increased spin waves excitation
 at higher temperature leads to $M_s \propto T^{3/2}$ \cite{spinwaves}.
 This is observed down to $T\leq 8K$, where the moment decreases below its expected value (Fig.13).

\subsection{Reduction of the magnetic moment at low temperature}

This low temperature moment reduction suggests a breakdown of spin collinearity. The drop from the expected value of the magnetization
 at low temperature gives a relative reduction $r=0.94-0.97$, 
and would yield an averaged canting angle $\theta_c$ of $\theta_c=arccos(r)\approx 14-20 deg$ \cite{brown}.
 Looking at the hysteresis loop at low temperature (Fig.14), another effect is observed simultaneously: the virgin magnetization curve for
 $T<8K$ changes notably. The shape of this curve recalls a pinning type magnet, but it lies outside the hysteresis loop boundary before saturation
 what is clearly unusual for a conventional magnet. We propose that the virgin magnetization
 reflects an irreversible metamagnetic transition, implying that the anisotropy is reinforced at low temperature. 
%The role of eddy currents in such a behavior has been emphasized long time ago,
% but it can not be proposed in our quasi-insulating sample ($R>10^6 \Omega$  for $T<250K$).
 %Recently, similar anomalous virgin curve has been observed was interpreted as a phenomenon of kinetic 
%arrest of the high temperature phase in materials presenting a first order magnetic transition \cite{sharma}.
 %As shown above, the sole magnetic transition which is clearly identified in this sample is continuous and such explanations can not be proposed here.
%Field induced ferromagnetism has been already reported for manganites compounds, where several experimental evidences
% point out the role of heterogeneous, spatially separated, magnetic phases. Even if the possibility of phase separation can
 %n%ot be definitively rejected, we do not observe experimental evidence of it. We propose that the virgin magnetization
 %reflects an irreversible metamagnetic transition. 
 In such a case, the curve represents the irreversible rotation of one sublattice of magnetization
 with antiferromagnetic ground state at a critical field $H_c$ (here $H_c\approx  0.45T$ at $T=2K$). 
Goodenough-Kanamori exchange rules indicate an antiferromagnetic superexchange pathways in the chains containing the $Fe^{II}$ ions.
 Metamagnetic-like behavior implies also competiting interactions and then ferromagnetic interactions (between next nearest neighbourg),
 which is consistent with the ferrimagnetism of the coumpound.
We observe a remarkable increase of $H_c$ between $T=5K$ and $T=2K$ (Fig.14), in the temperature range where the canting occurs.
It could arise from the reinforcement of the single ion magnetic anisotropy of $Fe^{II}$ ions as the temperature decreases.
Another possibility is an antisymmetric Dzyaloshinsky-Moriya (D-M) interaction which results from the anisotropic superexchange interaction \cite{dm},
and is allowed here because two neighboring $Fe^{II}$ ions are not related by a center of inversion.

\section{Conclusion}

We have measured and discussed the magnetic properties of an original mixed valency iron phosphate, focusing on the role of the complex structural disorder.
Its random distribution at a nanoscale is responsible for the slow dynamics of magnetic domain walls as evidenced in the frequency
 dependence of the dynamical susceptibility,
 but does not perturb significantly the long-range ferrimagnetic ordering, which is found of 3D Heisenberg type. At low temperature,
 the breakdown of the Bloch law indicates a spin canting.
 This implies a reinforced anisotropy at low temperature, which is likely responsible for the irreversible metamagnetic transition which appears
 in the same temperature range. 
This study shows how the role of disorder can be important, but can have also contrasted consequences, for magnetic properties.

\newpage
\begin{figure}[t!]
\begin{center}
\includegraphics*[width=8.0cm]{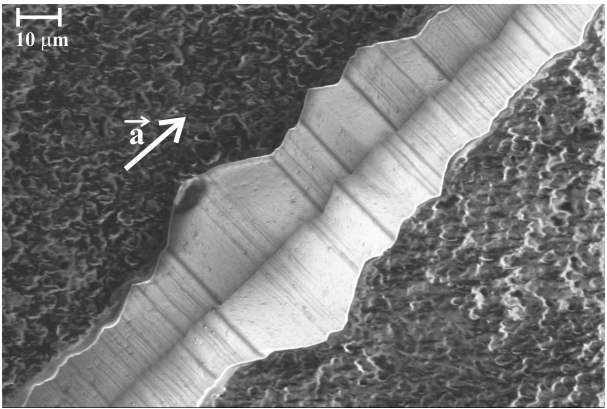}
\end{center}
\caption{MEB picture of a needle-like crystallite of $\beta -(NH_4)Fe_2(PO_4)_2$, with the direction of the a-axis.}
\label{fig.1}
\end{figure}

\begin{figure}[t!]
\begin{center}
\includegraphics*[width=8.0cm]{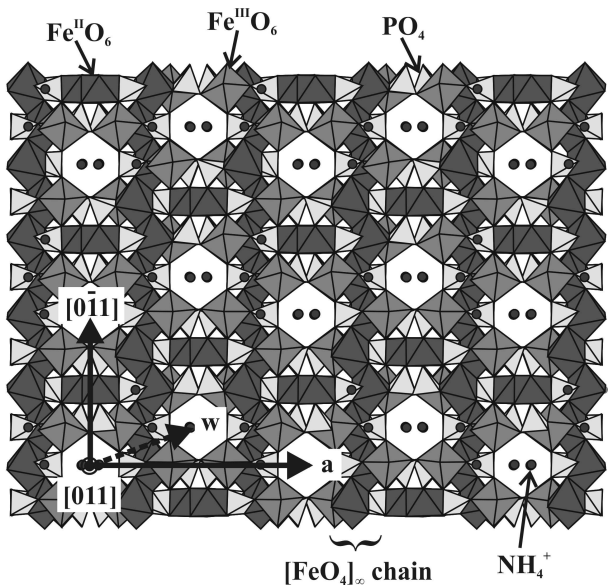}
\end{center}
\caption{Projection of the $\beta-(NH_4)Fe_2(PO_4)_2$ structure along [011].}
\label{fig.2}
\end{figure}

\begin{figure}[t!]
\begin{center}
\includegraphics*[width=8.0cm]{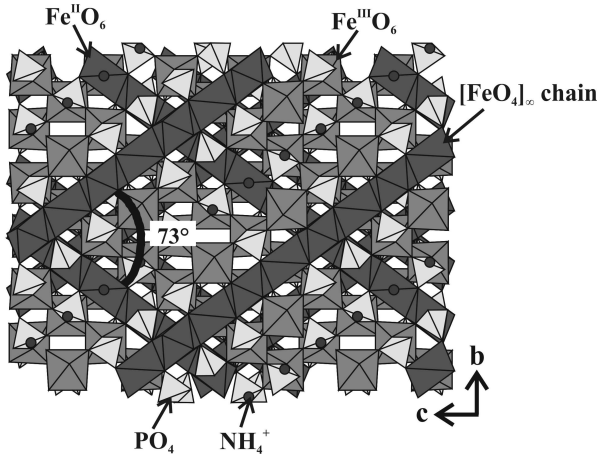}
\end{center}
\caption{ Projection of the $\beta-(NH_4)Fe_2(PO_4)_2$ structure along  the a-axis (i.e. perpendicularly to the (bc) plan).}
\label{fig.3}
\end{figure}

\begin{figure}
\includegraphics[width=0.48\textwidth]{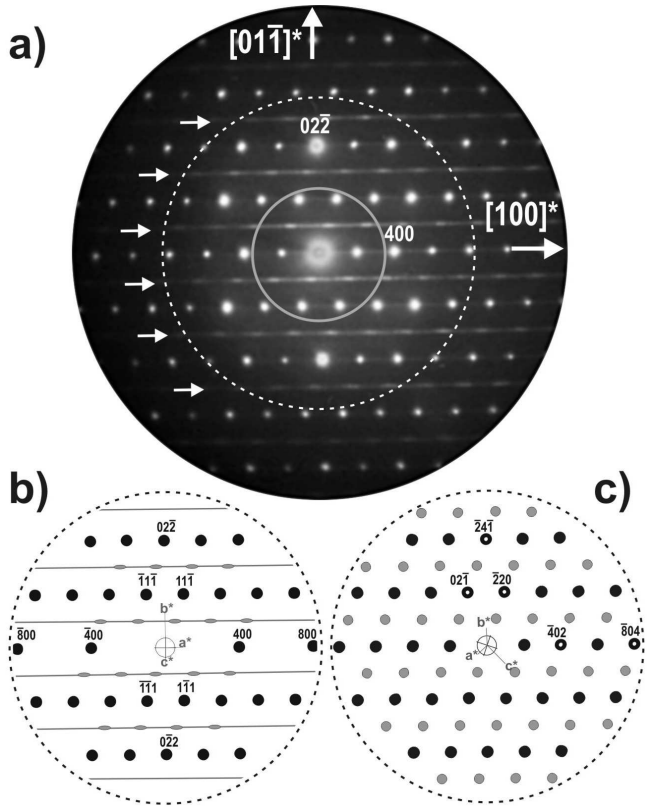}
\caption{a) [011] SAED zone axis patterns observed on a $\beta$-(NH$_4$)Fe$_2$(PO$_4$)$_2$ crystallite.
 It can be described as the regular alternation along [01$\bar{\mathrm{1}}$]* of two type of rows parallel to [100]* one
 being diffuse streaks at d$^{*}$=o$\cdot\frac{1}{2}\cdot$d$^{*}_{01\bar{1}}$ (o for odd integer) and the other made of spots
 at d$^{*}$=n$\cdot$d$^{*}_{01\bar{1}}$. b) Schematic representation of the dashed circular area in a) and where the spots
 (in dark) related to the average structure are indexed considering the $F_{ddd}$ orthorhombic cell. In gray is represented the
 diffuse streaks where intensity reinforcement can be observed. c) The diffuse nodes can actually be indexed consideringa [112] ZAP of the
 $\alpha$-(NH$_4$)Fe$_2$(PO$_4$)$_2$ ordered monoclinic form \cite{alpha-sosso}.}
%At positions corresponding to d*=$\frac{1}{2}$d*$_{01\bar{1}}$
\label{fig.4}
\end{figure}

\begin{figure}
\includegraphics[width=0.48\textwidth]{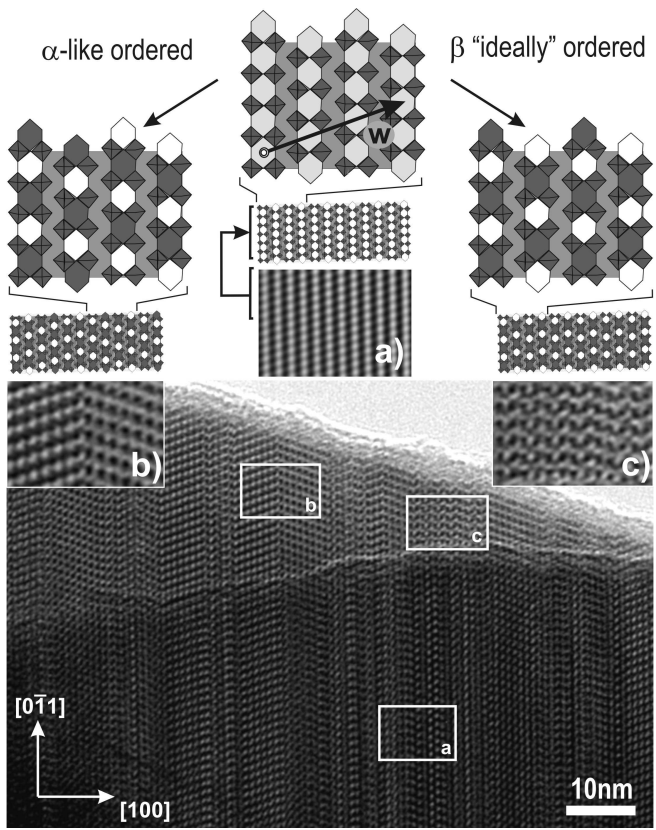}
\caption{$\sim$5\AA\ resolution TEM image obtained from a $\beta$-(NH$_4$)Fe$_2$(PO$_4$)$_2$ crystallite. 
%with the presence of twinning along the [100] direction. 
a) Fourier filtered image generated by selecting only the spots present in the diffractogram observed in the Fourier
 transform of the a area. Such an image allow to locate the hexagonal tunnels present in the structure regardless how
 they are filled as illustrated in the schematic structure drawing above the image. 
Depending on the filling sequence along {\bf w}$\sim$[2$\bar{1}$1] between NH$_4^+$ groups (white hexagon) and
 [FeO$_{4}$]$_{\infty}$ chains (dark grey hexagon), it is possible to generate either a $\alpha$- or $\beta$-type ordering.
 b) enlargement (x2) of the experimental image (b area) showing that this part of the crystal can be seen as a
 $\alpha$-(NH$_4$)Fe$_2$(PO$_4$)$_2$ twin structure at a nanoscale. c) enlargement (x2) of the experimental image
 (c area) showing that locally this part of the crystal can be seen as an "ideally" $\beta$-(NH$_4$)Fe$_2$(PO$_4$)$_2$ ordered structure.}
\label{fig.5}
\end{figure}

\begin{figure}[t!]
\begin{center}
\includegraphics*[width=8.0cm]{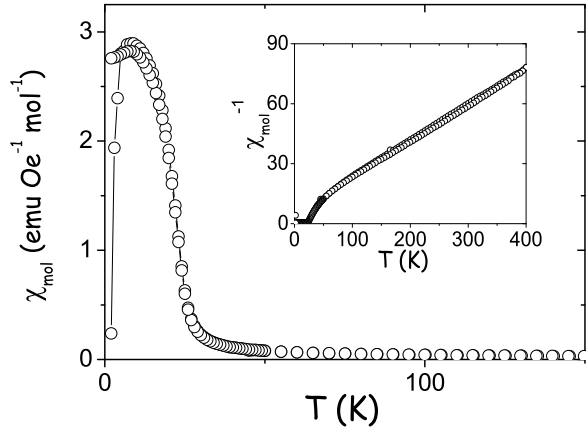}
\end{center}
\caption{Molar magnetic susceptibility as function of the temperature measured during field cooling and zero field cooling procedures with an applied field of 0.1T.
In the inset is shown the inverse molar susceptibility to evidence the Curie-Weiss regime at high temperature.}
\label{fig.6}
\end{figure}

\begin{figure}[t!]
\begin{center}
\includegraphics*[width=8.0cm]{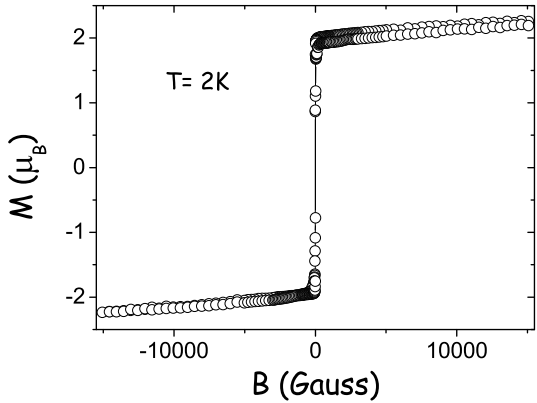}
\end{center}
\caption{Magnetic cycle after saturation of the magnetization measured for a temperature $T=2K$. The magnetization is per formula unit.}
\label{fig.7}
\end{figure}

\begin{figure}[t!]
\begin{center}
\includegraphics*[width=8.0cm]{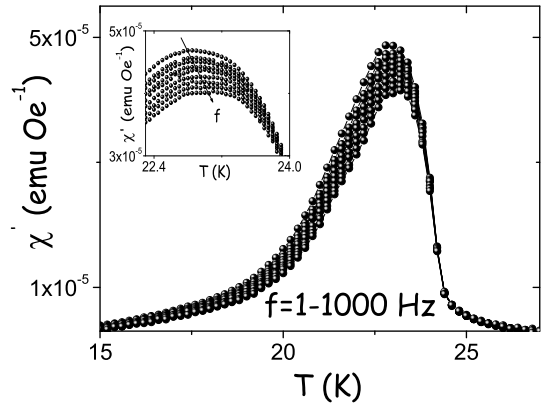}
\end{center}
\caption{Ac susceptibility for $H=0$ and for different frequencies $f=1, 5, 10, 25, 50, 120, 250, 500$ and $1000 Hz$. In the inset are shown similar measurements, taken with small temperature steps.}
\label{fig.8}
\end{figure}

%\begin{figure}[t!]
%\begin{center}
%\includegraphics*[width=8.0cm]{freq.eps}
%\end{center}
%\caption{Frequency as function of the temperature of the maximum of $\chi^{'}$ in a log-linear scale, deduced from the data of the fig.8.}
%\label{fig.9}
%\end{figure}

\begin{figure}[t!]
\begin{center}
\includegraphics*[width=8.0cm]{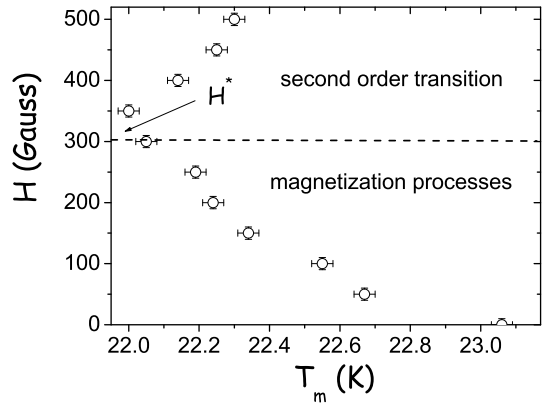}
\end{center}
\caption{Applied magnetic field as function of the temperature of the maximum of suceptibility ($f=7 Hz$).
 Note the inversion of the dependence at $H=H^*\sim 300 G$.}
\label{fig.9}
\end{figure}

\begin{figure}[t!]
\begin{center}
\includegraphics*[width=8.0cm]{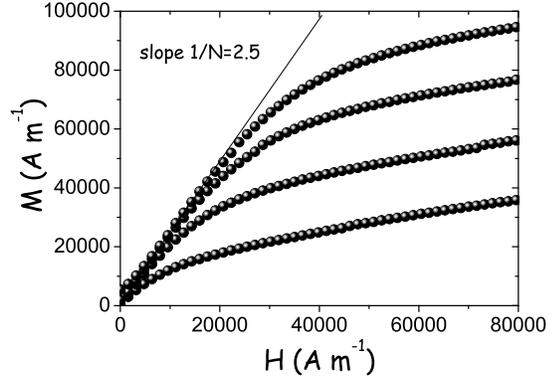}
\end{center}
\caption{Magnetization as function of the increasing magnetic field, for temperatures $T=23,22,21$ and $20 K$ from bottom to top.
 The straight line is the line of demagnetizing field up to $H\sim 20000 A/m\sim 250 G$ at 20K.}
\label{fig.10}
\end{figure}

\begin{figure}[t!]
\begin{center}
\includegraphics*[width=8.0cm]{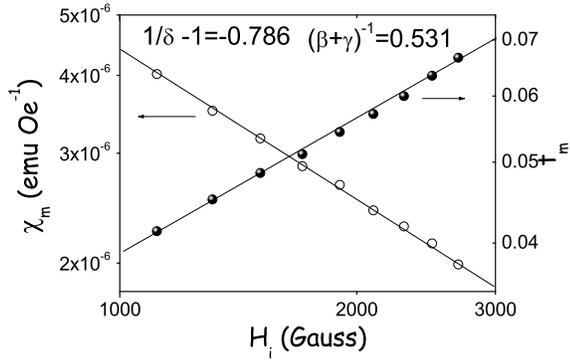}
\end{center}
\caption{Maximum of the in phase susceptibility ($f=7 Hz$) as function
 of the internal magnetic field $H_i$(corrected from the demagnetizing field) in log-log scale.
Also shown is the reduced critical temperature as function of $H_i$ in log-log scale.}
\label{fig.11}
\end{figure}

\begin{figure}[t!]
\begin{center}
\includegraphics*[width=8.0cm]{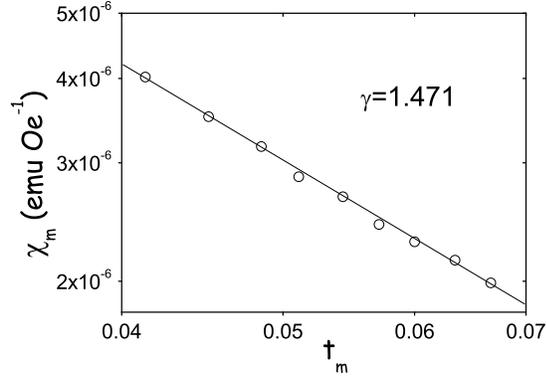}
\end{center}
\caption{Maximum of the in phase susceptibility ($f=7 Hz$) as function of the reduced temperature in log-log scale.}
\label{fig.12}
\end{figure}

\begin{figure}[t!]
\begin{center}
\includegraphics*[width=8.0cm]{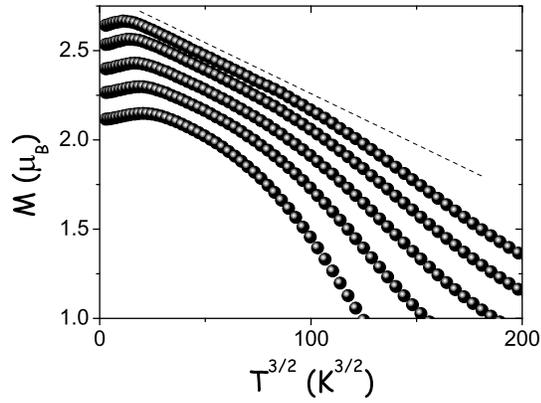}
\end{center}
\caption{Field cooled magnetization as function of $T^{3/2}$, for fields $H=1, 2,3, 4$ and $5T$ from bottom to top.
 The dotted line corresponds to the temperature variation of the Bloch law which becomes apparent with the increase of the applied field.
 Note the decrease of the measured magnetisation at low temperature, compared to the dotted line. The magnetization is per formula unit.}
\label{fig.13}
\end{figure}

\begin{figure}[t!]
\begin{center}
\includegraphics*[width=8.0cm]{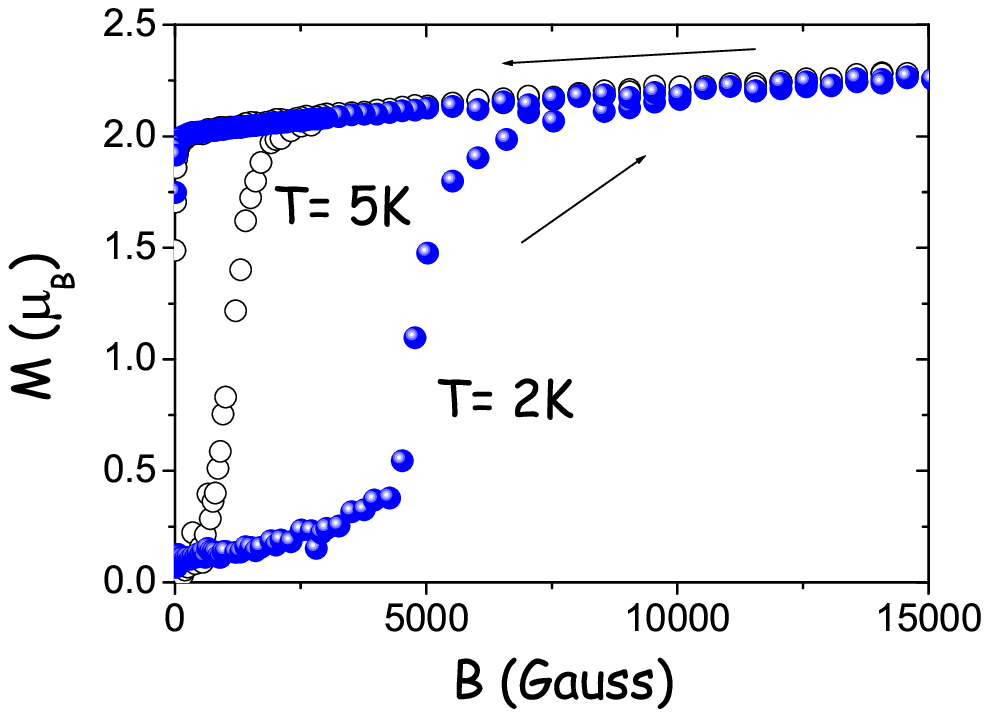}
\end{center}
\caption{(Color Online) First magnetization curve as function of the increasing and then decreasing magnetic field, for temperatures $T=5$ and $2 K$, showing an irreversible metamagnetic transition. The magnetization is per formula unit.}
\label{fig.14}
\end{figure}

\begin{references}
\bibitem{notebene}The $(H,T_m)$ points for $H<H^*$ vary as $(T-T_c)^{3/2}$ where
$T_c$ is the temperature of the zero field suceptibility maximum. This dependence is similar of the Almeida-Thouless
 line of spin glasses \cite{AT}, but as other spinglasses properties are not verified, it is only incidental.
\end{references}
\end{document}